\documentclass[twoside,twocolumn,10pt]{article}
\usepackage{lipsum}  
\usepackage{fancyhdr}
\usepackage{geometry}
\usepackage{multicol}
\usepackage{abstract}
\usepackage{graphicx}
\usepackage{caption}
\usepackage{titlesec}
\usepackage{ragged2e}  
\usepackage{amssymb}  
\usepackage{parskip}  
\usepackage{amsmath}  
\usepackage{newtxtext}  
\usepackage{booktabs}  
\usepackage{array}     
\usepackage{subcaption}

\newcolumntype{L}[1]{>{\raggedright\arraybackslash}p{#1}} 
\newcolumntype{C}[1]{>{\centering\arraybackslash}p{#1}}   
\newcolumntype{R}[1]{>{\raggedleft\arraybackslash}p{#1}}  

\usepackage[font=footnotesize, labelfont=bf, justification=raggedright, format=plain]{caption}

\usepackage[backend=bibtex,style=ieee]{biblatex}
\defbibheading{bibliography}[\refname]{%
  \section{\MakeUppercase{REFERENCES}}}

\titleformat{\section}{\large\bfseries\uppercase}{\thesection.}{1em}{}

\titleformat{\subsection}{\normalfont\bfseries\flushleft}{\thesubsection.}{1em}{}

\titleformat{\subsubsection}{\normalfont\bfseries\itshape\flushleft}{\thesubsubsection}{1em}{}

\newcommand{\keywordsname}{Keywords}

\fancypagestyle{firstpage}{
  \fancyhf{}  
  \fancyhead[LO]{\small August 2026}  
  \fancyhead[RO]{\small arXiv}  
  \fancyfoot[L]{\footnotesize Submitted: August 2026\\\textsuperscript{*}Corresponding Author: 
  \textless miao.guo@kcl.ac.uk\textgreater}
   \fancyfoot[R]{\footnotesize \copyright\ 2026 by the Author(s).}
}

\title{\fontsize{16}{20}\selectfont\textbf{PyOMES: an open-source framework for biochemical process modelling}}
\author{
    \fontsize{12}{14}\selectfont
    Ethan Errington, Tom Vinestock, Jaewook Lee, Miao Guo \textsuperscript{*}\\[1ex]
    \fontsize{12}{14}\selectfont
    Department of Engineering, King's College London, London, UK \\
}
\date{}

\begin{document}

\twocolumn[
\begin{@twocolumnfalse}
\vspace{-0.5cm}  
\raggedright

{\fontsize{11}{13}\selectfont{Research Article}}
\vspace{-0.8cm}  

\maketitle

\thispagestyle{firstpage} 
\begin{abstract}
\fontsize{11}{13}\selectfont
\vspace{1em}
\justifying
PyOMES is a Python-based, Open-source Modelling Environment for (bio)chemical process Simulation that aims to simplify the modelling of dynamic (including steady state) processes. This is done in a generalied, modular way to facilitate modelling a broad range of biological, chemical, and biochemical systems under a single modelling framework. PyOMES has been built to be accessible to a broad range of users - ranging from those with little modelling experience, such as experimentalists and students, through to more experienced power-users. Here, an introduction is provided to the PyOMES software including a summary of the design, architecture and vision. Use cases are then provided to demonstrate applicability of PyOMES to a number of (bio)chemical process modelling scenarios. 
Comparisons of predictions against existing benchmark software (i.e. PHREEQC) demonstrate the robustness of the package. Finally, a summary is provided of future directions for the PyOMES package - highlighting its establishment as a unified modelling framework and its potential for community-driven improvement as future developments.

\vspace{-0.1cm}
\noindent\textbf{Keywords}
 bioprocess modelling, process systems engineering, open-source software, modelling framework, PyOMES.
 
\end{abstract}

\vspace{1cm}
\end{@twocolumnfalse}
]
\section{Introduction}
\fontsize{12}{14}\selectfont
\par Our understanding of biological and chemical – “(bio)chemical" - systems has developed significantly in recent decades. This has been led by the contributions of experimental and theoretical lines of inquiry summed across many different scientific and engineering disciplines. Paired with recent advancements in the power and availability of computational resources, this now enables us to predict and control many important properties of (bio)chemical processes.
\par Despite this, computational modelling of (bio)chemical processes still faces a significant structural challenge. Specifically, the expertise needed to simulate (bio)chemical processes exist at the intersection of many different scientific fields, while the software infrastructure needed to allow researchers to integrate these parts into a cohesive whole is lacking \cite{Helikar2021}. The result is a constraint on the number of practising (bio)chemical experimentalists that are capable of using computational modelling methods to inform and enhance the productivity of research and development pipelines.

\par If this challenge can be addressed, (bio)chemical simulations can be used more widely, by a wider range of practitioners, and in closer dialogue with experimental approaches, to enhance the quality and rate of R\&D. For example, there exists a broad range of literature demonstrating the benefits of simulation-capable models in areas such as model based design of experiment \cite{Geremia2026ModelBasedDesign}, model based predictive control \cite{Rawlings2026ModelPredictiveControl}, and digital twins \cite{Rebello2025DigitalTwins}.

\par In many cases, the development of tailored, user-centred modelling software can greatly help experimentalists incorporate (bio)chemical simulations into their research. In principle, all this requires is software available in a format that is easy to understand, easy to access, and easy to use. However, establishing a package of this type is difficult exactly because of the challenge posed by covering several separable conceptual modelling areas intrinsic to handling dynamic and steady state (bio)chemical systems. These include understanding of mass \& energy balances, thermodynamics, control theory, mass transport phenomena,  mathematical model formulation, and the solution of systems of differential equations.


\par Accessible, integrated modelling tools remain a barrier to the wider adoption of modelling, particularly among experimentalists. As models increase in complexity, spreadsheet- and script-based implementations can become difficult to maintain, debug and reuse. Similar arguments regarding the limitations of Excel and script based modelling have been discussed in the fields of bioinformatics \cite{Djaffardjy2023} and health economics \cite{Baio2016}. Within biochemical process modelling, these barriers discourage the development of coupled dynamic models that simultaneously represent the core chemical/biological kinetics of interest as well as environmental and transport phenomena. Variables such as pH and gas transfer are important in many biological and chemical processes \cite{GonzlezFigueredo2019}, but are under-modelled by experimentalists due to the limitations of existing modelling software.
\par In this work, and in response to the previously mentioned problems, we introduce PyOMES - a Python-based, Open-source Modelling Environment for process Simulation - enabling 
simulation of dynamic (including steady-state) (bio)chemical processes. First, the code architecture and design ideas underlying the PyOMEs modelling framework are introduced, and then three separate use-case scenarios demonstrating PyOMES' functionality are provided.

\section{Package Overview}

\begin{figure*}[!h]
     \centering
     \includegraphics[width=\linewidth]{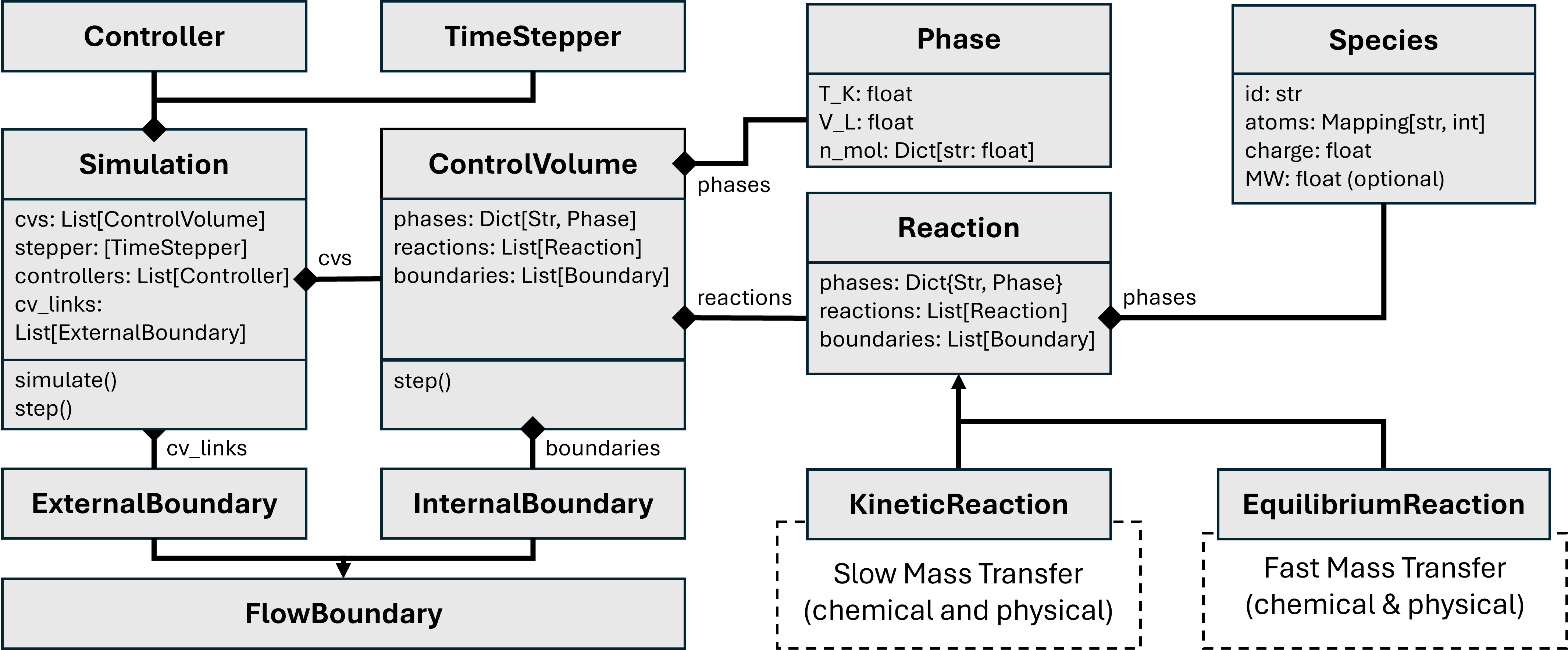}
     \caption{A simplified UML Class Diagram describing the core classes present in the PyOMES python package.}
     \label{fig:class diagram}
 \end{figure*}
\subsection{User Centered Design}
\fontsize{12}{14}\selectfont
PyOMES is designed with three main user types in mind: 
\begin{enumerate}
    \item \textit{Power-users} - Users with a strong modelling background and knowledge of the PyOMES architecture. For these users, the modular design of PyOMES enables significant customizability for developing complex models based on the use of \textit{Protocols} (see Section \ref{sec: Modular design}) in the package architecture.
    \item \textit{Model-aware users} - Users familiar with the main modelling concepts relevant to their systems of interest. These users will find that the architecture of PyOMES enables them to specify virtually all standard modelling use cases to high levels without the need to deeply understand the PyOMES architecture or rely on restrictive supporting functions.
    \item \textit{Novice-users} - Users with minimal experience in modelling (bio)chemical systems (including pure experimentalists and students). For these, PyOMES provides supporting functions that enable rapid development of common model concepts without need for detailed package understanding. This includes shortcut functions for common kinetic models, thermodynamic database parameters (see Section \ref{sec: thermodynamics framework}), and ODE solvers (see Section \ref{sec: solving differential equations}).
\end{enumerate}

\subsection{Package Architecture}
\fontsize{12}{14}\selectfont
PyOMES is an installable python package written in the Python programming language \cite{python310}. It allows users to design and simulate dynamic (including steady-state) (bio)chemical processes using an object-oriented programming approach.
\par The architecture of PyOMES relies on the description of a (bio)chemical process according to a \textit{Simulation} object which handles running simulations using a control volume approach (see  Section \ref{ref:cv approach}) as shown in Figure \ref{fig:class diagram}.

\par The \textit{Simulation} object contains most of the information associated with the computational model. The \textit{ControlVolume} object acts as a central building block for material characteristics and properties (see also Section \ref{ref:cv approach}). The \textit{Phase} object acts as a source of truth for thermodynamic and state accounting within a given \textit{ControlVolume}. The Species object describes the chemical species present in any model (including those within biological organisms, to ensure mass balance and reaction stoichiometries are respected). Finally, the \textit{Reaction} objects act to describe transient phenomena that can occur within a given \textit{ControlVolume} object. Importantly, the \textit{Reaction} type  is considered to include both chemical reactions (including bond breakage and formation) as well as mass exchange phenomena occurring within a single \textit{ControlVolume} (e.g. gas-liquid mass exchange).

\subsection{Modular Design Principle}\label{sec: Modular design}
PyOMES has been developed with the vision of enabling modelling in a way that is as generalisable and modular as possible. This allows complex models to be developed from the aggregation of a number of relatively simple models, improving human readability. In doing so, the design approach can be summarised in terms of two major modular design axes: 1) conceptual modularity (i.e. the separation of conceptually distinct phenomena within any model), and 2) spatial modularity (i.e. separation of physical locations within a model using the \textit{ControlVolume} paradigm). 

\par \textbf{Conceptual Modularity}: This is achieved through use of the \textit{Protocol} functionality available within the python programming language.  Use of the \textit{Protocol} approach allows PyOMES to define a generalised, minimal architectural skeleton describing ways in which distinct modelling areas must interact, process, and pass information. This allows PyOMES to ship as a package with a small number of well-tested default implementations for common modelling use-cases while allowing power-users to develop advanced modelling scenarios either using their own custom code or other imported packages and software.

\par \textbf{Spatial Modularity}\label{ref:cv approach}: The use of a control volume paradigm is central to the modelling approach of the PyOMES package and its modelling framework. In PyOMES, \textit{ControlVolume} objects represent and decompose as distinct units in physical space with predefined characteristics ( e.g. molecular states, (bio)chemical properties and phases). In this way, the definition of any process model present in PyOMES is then describable as the phenomena arising from instantiation of one or more control volumes with pre-defined characteristics.  

\par The control-volume abstraction benefits PyOMES independently of how many compartments a model uses. Each \textit{ControlVolume} object acts as a self-contained unit against which mass, charge, and element conservation can be checked automatically, and a module in which mass transport phenomena can be described topologically without reference to the rest of the system (see Section \ref{sec: Mass Transport Phenomena}). Additionally, the \textit{Simulation}-\textit{ControlVolume} contract enables the scaling of models to multi-compartment systems by composing several control volumes through transport links - e.g. digester trains, chromatography columns, stratified fermenters.

\subsection{Thermodynamic Framework}\label{sec: thermodynamics framework}
 PyOMES represents the thermodynamic assumptions of a model through a dedicated object called the \textit{ThermoFramework}. This object defines the thermodynamic model used to calculate quantities such as activities, activity coefficients, and equilibrium relationships, and acts as a shared source of thermodynamic information for the other layers of the package. This keeps the thermodynamic assumption of a model as a single, swappable choice that is adhered to across a PyOMES model. Crucially, \textit{ThermoFramework} follows the same protocol-based extensibility used elsewhere in the framework (see Section \ref{sec: Modular design}) to enable users to easily supply custom models that satisfy their modelling needs.

\subsection{Mass Transport Phenomena}\label{sec: Mass Transport Phenomena}
The approach to modelling mass exchange in PyOMES is decomposed into three conceptual axes - physiochemical, spatial and temporal. 

\par \textbf{Physiochemical Decomposition}: This distinguishes between physical transport and chemical mass exchange. Chemical transport involves the making and breaking of bonds -  meaning a change in tracked mass between distinct chemical states (e.g. within \textit{Phase} objects). Physical transport involves the movement of material in space without chemical changes (e.g. across \textit{Phase} objects).

\par \textbf{Spatial Decomposition}: Here, a distinction is made between mass transport occurring within a single \textit{ControlVolume} object (intra-CV) and that occurring across \textit{ControlVolume} objects (inter-CV). This dimension may be thought of as the spatial dimension of mass transport phenomena. 

\par \textbf{Temporal Decomposition}: This distinguishes mass transport (based on the timescales over which they occur) as being either dynamic or instantaneous. Dynamic transport represents phenomena requiring time-resolved kinetic modelling through the use of ordinary differential equations (ODEs). Introduction of these phenomena require dedicated solvers to handle the logistics of ODE integration.
Instantaneous transport represents phenomena that occur relatively quickly in the model; such cases are described using simple algebraic equations. Options present within the package allow for easy use of SciPy implementations \cite{2020SciPy}. 

\subsection{Solution of Differential Equations}\label{sec: solving differential equations}
PyOMES advances a simulation including kinetic phenomena using a two-axis solver architecture that separates how each control volume integrates its own local physics from how the whole multi-compartment system advances together.
\par \textbf{Intra-\textit{ControlVolume}}:  The \textit{StepSolver} protocol enables assignment of specific algorithms for solving differential equations to individual \textit{ControlVolume} objects. This provides a simple basis for running single \textit{ControlVolume} model objects as well as running multi-compartmental simulations in which the importance of inter-\textit{ControlVolume} transport is negligible.
\par \textbf{Inter-\textit{ControlVolume}}: The \textit{SystemSolver} protocol enables description of more solver strategies accounting for scenarios including flows  across multiple \textit{ControlVolume} objects in multi-compartment models. This axis is particularly important in handling areas such as differential algebraic equations (DAEs), important when modelling both quick equilibrium reactions and slower kinetic reactions within the same schema.

 
\section{Modelling Use Cases}
\begin{figure*}[h!]
    \centering
    \includegraphics[width=0.25\linewidth]{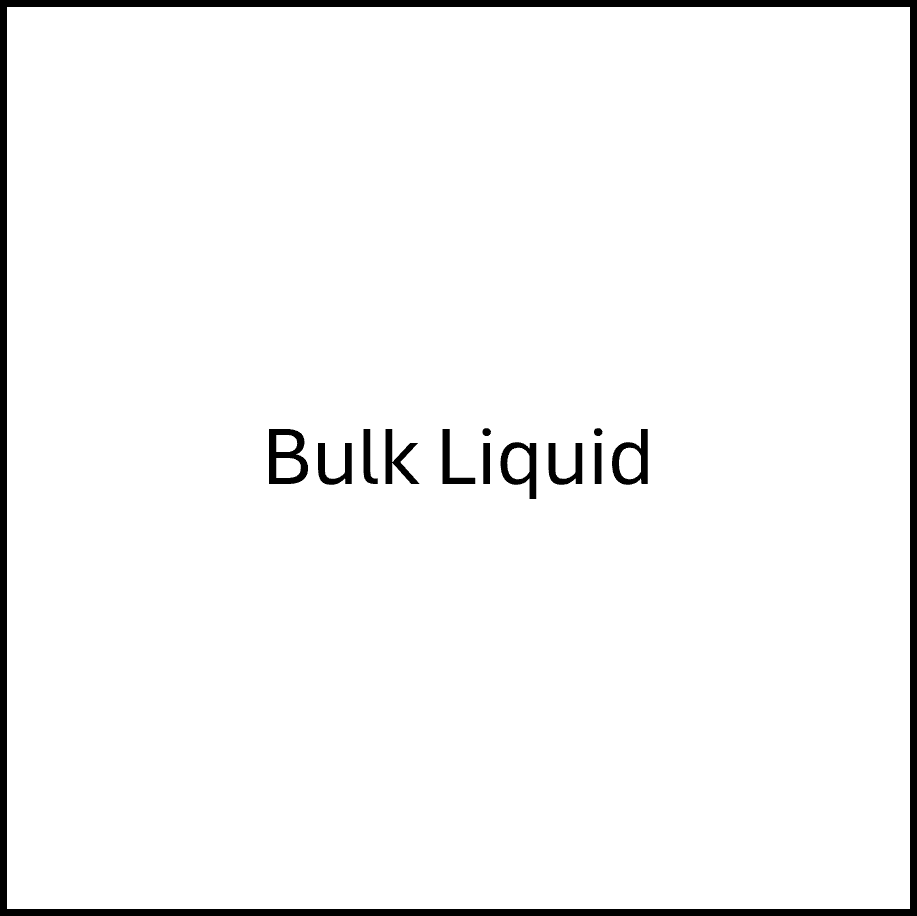}
    \hfill
    \includegraphics[width=0.25\linewidth]{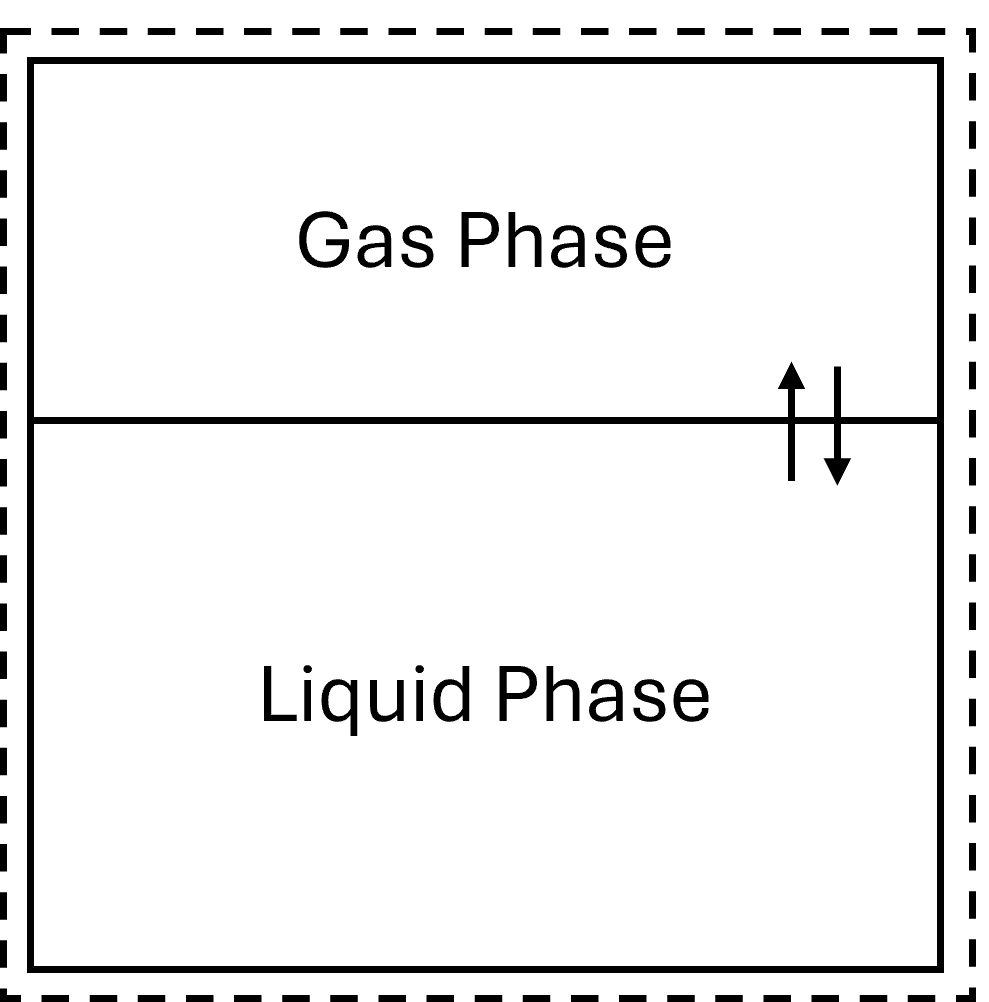}
    \hfill 
    \includegraphics[width=0.35\linewidth]{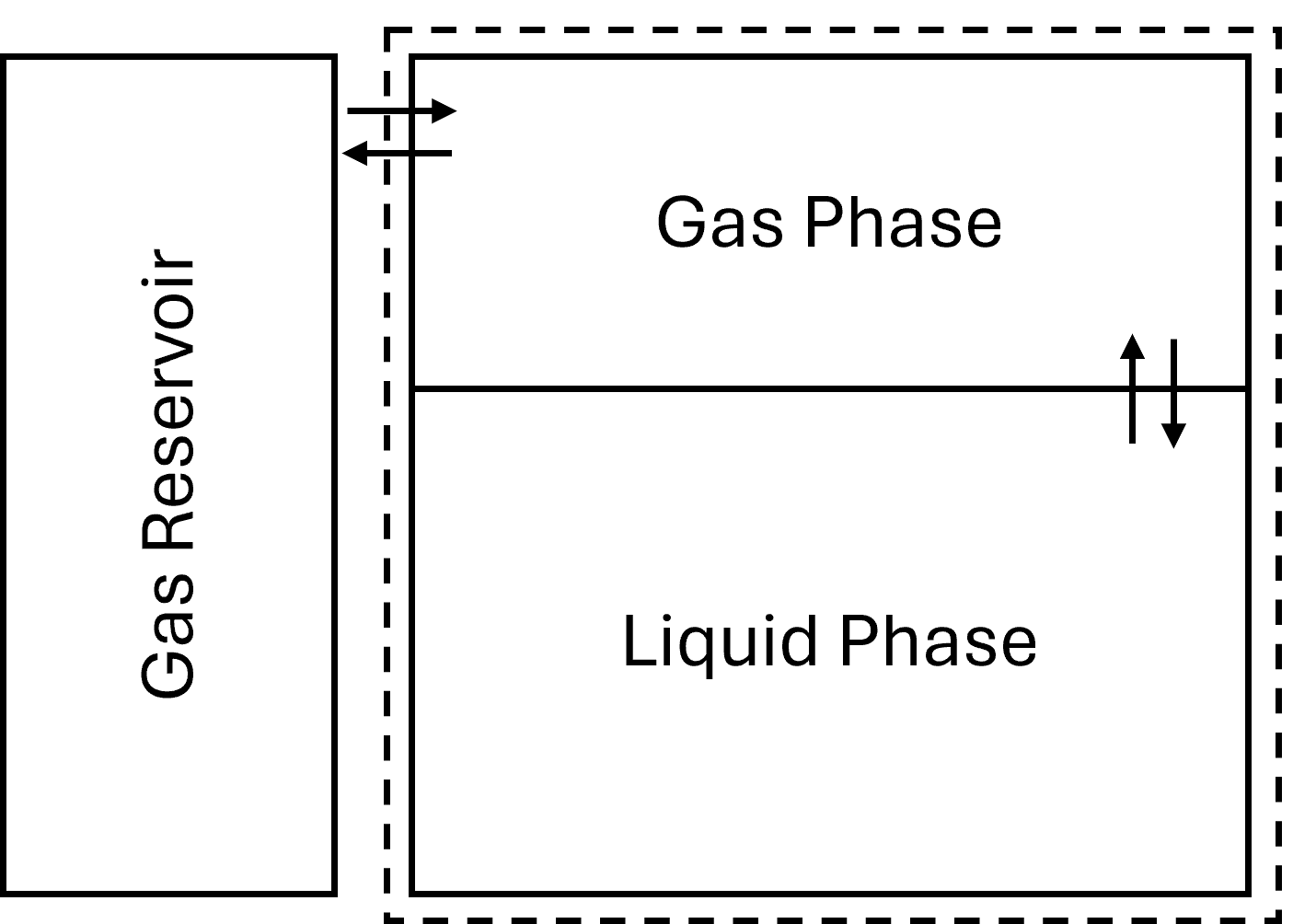}
    \caption{Conceptual schematics for the three use-cases presented in this work - equilibrium pH of an aqueous solution in a closed system (left), dynamic equilibration of CO$_2$ in a gas-liquid system connected to the atmosphere via a gas-permeable membrane (middle), and steady-state aerobic fermentation of E.Coli in a continuous stirred tank reactor (right).}
    \label{fig: modelling use case schematics}
\end{figure*}

\fontsize{12}{14}\selectfont
Three use-cases of the PyOMES package will be presented to demonstrate functionality of key object types and their relation to standard (bio)chemical simulation. Each use case is developed with a focus of demonstrating how PyOMES can be used to answer different kinds of experimental research questions through the use of computational simulation. We start with a simple use case, and gradually build up the complexity of the modelled system to provide a gentle introduction to the packages. The use cases are visualised in Figure \ref{fig: modelling use case schematics}. 

\subsection{Aqueous pH Prediction}\label{sec: pH prediction methods}
This use case models the relationship between composition and pH for an aqueous solution.
\par The compositions considered are for mixtures of potassium dihydrogen phosphate (KH$_2$PO$_4$) and ammonium chloride (NH$_4$Cl) in the range of 0 to 100 mmol/L at a temperature of 25 C. These are equivalent to mass concentrations of 0 to 13.60 g L$^{-1}$ KH$_2$PO$_4$ and 0 to 5.35 g L$^{-1}$ NH$_4$Cl respectively.

\par As part of the model it is assumed that both salts have already dissolved and dissociated into their constituent ions (Equations \ref{eq: ammonium chloride dissolution} and \ref{eq: potassium phosphate dissolution}), and that solution equilibria are limited to water autoionisation (Equation \ref{eq: water dissociation}), ammonium ion dissociation (Equation \ref{eq: ammonium ion dissociation}), and the various states of phosphate protonation (Equation \ref{eq: phosphate dissociation 1} to \ref{eq: phosphate dissociation 3}). 

\begin{equation}\label{eq: ammonium chloride dissolution}
    NH_4Cl(s) \rightarrow{} NH_4^+(aq) + Cl^-(aq)
\end{equation}
\begin{equation}\label{eq: potassium phosphate dissolution}
    KH_2PO_4(s) \rightarrow{} K^+(aq) + H_2PO_4^-(aq)
\end{equation}
\begin{equation}\label{eq: water dissociation}
    H_2O(l) \rightleftharpoons H^+(aq) + OH^-(aq)
\end{equation}
\begin{equation}\label{eq: ammonium ion dissociation}
    NH_4^+(aq) \rightleftharpoons  NH_3(aq) + H^+(aq)
\end{equation}
\begin{equation}\label{eq: phosphate dissociation 1}
    H_3PO_4(aq) \rightleftharpoons H_2PO_4^{-}(aq) + H^+(aq)
\end{equation}
\begin{equation}\label{eq: phosphate dissociation 2}
    H_2PO_4^-(aq) \rightleftharpoons HPO_4^{2-}(aq) + H^+(aq)
\end{equation}
\begin{equation}\label{eq: phosphate dissociation 3}
    HPO_4^{2-}(aq) \rightleftharpoons PO_4^{3-}(aq) + H^+(aq)
\end{equation}

\par All equilibrium equations (Equations \ref{eq: water dissociation} to \ref{eq: phosphate dissociation 3}) are assumed to occur instantaneously. The Davies model \cite{Davies1962IonAssociation} was used to account for the effect of the ionic strength of solution on solution non-ideality. Solution of calculations necessary to identify equilibrium pH was carried out using a Newton-Raphson based solver available within the PyOMES package based on a standard log$K$-based problem formulation. To establish reliability of predictions made by PyOMES, a number of predictions were also benchmarked against predictions made for the same system by the PHREEQC \cite{parkhurst2013phreeqc} software. For further information, see the Supporting Information.


\subsection{Gas-Liquid Equilibration}\label{sec: gas-liquid methods}
\par This use case models the relationship between pH and time for (initially) pure water that remains in contact with atmospheric air.

The composition (molar basis) of atmospheric air is assumed to be 20.946\% oxygen (O$_2$), 78.084\% nitrogen (N$_2$) and 400 ppm carbon dioxide (CO$_2$).  All phases are assumed to maintain a constant temperature and pressure of 25$^\circ$C and 1 atm, respectively. The air volume is large relative to the liquid volume ($V_G = 1000$ L, $V_L = 0.1$ L), so gas-phase depletion was negligible. Under this pseudo-infinite-headspace assumption, the liquid concentration and pH trajectories are governed primarily by the volume-normalised mass-transfer
coefficient $k_La$ (see Equation \ref{eq: gas-liquid kinetics}).

\par As part of the model, the transfer of material (N$_2$, O$_2$, and CO$_2$) across the gas-liquid interface is assumed to occur according to the stoichiometry described in Equations \ref{eq: O2 vle} to \ref{eq: co2 vle}.

\begin{equation}\label{eq: O2 vle}
    O_2(g) \rightleftharpoons O_2(aq)
\end{equation}
\begin{equation}\label{eq: n2 vle}
    N_2(g) \rightleftharpoons N_2(aq)
\end{equation}
\begin{equation}\label{eq: co2 vle}
    CO_2(g) \rightleftharpoons CO_2(aq)
\end{equation}
\par Once in the aqueous phase, dissolved CO$_2$ was modelled using a lumped carbonate equilibrium, in which hydrated carbonic acid is not represented as a separate species. The carbonate speciation was described by:
\begin{equation}\label{eq: HCO3 formation}
    CO_{2,(aq)} + H_2O \rightleftharpoons{} H^+ + HCO_3^{-}(aq)
\end{equation}
\begin{equation}\label{eq: CO3 formation}
    HCO_3^{-}(aq) \rightleftharpoons{} CO_3^{2-}(aq) + H^+
\end{equation}
\par As before, all aqueous equilibrium equations were assumed to occur instantaneously and solved as previously described (Section \ref{sec: pH prediction methods}). In contrast, transfer across the gas-liquid interface was modelled following the first-order rate process described in Equation \ref{eq: gas-liquid kinetics} 
\begin{equation}\label{eq: gas-liquid kinetics}
    \frac{dn_{L,i}}{dt} \frac{1}{V_L} = -\frac{dn_{G,i}}{dt}\frac{1}{V_L} = k_{L}a(C_{L,i}^*-C_{L,i})
\end{equation}
where $n$ is the moles of substance $i$ (mol), $V_l$ is the liquid volume (L) $k_La$ is the volumetric gas transfer coefficient (h$^{-1}$), $C_{L,i}^*$ is the equilibrium concentration of substance $i$ (mol L$^{-1}$), and $C_{L,i}$ is the current concentration of the substance in solution (mol L$^{-1}$) and $L$ and $G$ represent the liquid and gaseous phases respectively. Note that $k_La$ is the product of two terms, $k_L$ and $a$. $k_L$ is a kinetic constant for transfer of substance from the gas phase to the liquid phase (m hr$^{-1}$) and $a$ is the volume-normalised interfacial area of the gas liquid interface (m$^2$ m$^{-3}$). These terms can be hard to separate, so are often left together as a lumped parameter.

\par First, to determine the effect of this gas transfer on the pH of water that is initially pure, a constant mass transfer coefficient ($k_La$) of 96 h$^{-1}$ was assumed for each gas (N$_2$, O$_2$, and CO$_2$), based on \cite{Vinestock2026FusariumFermentation}. To solve the ODE associated with mass transfer (Equation \ref{eq: gas-liquid kinetics}) the SciPy \cite{2020SciPy} implementation of the backward differentiation formula (BDF) solver was used. 

\par Finally, the effect of $k_La$ on effective time to reach approximate equilibration was investigated by running a set of simulations for $k_La$ varying in the range of 0.1 to 1000 h$^{-1}$. For this, the point of effective equilibration was defined as the time at which the aqueous concentration of CO$_2$ reached 95\% of its equilibrium value.

\subsection{Aerobic Biomass Fermentation}
\par This use case models the aerobic growth of a single-celled organism (\textit{E. coli}) within a gas-sparged continuous stirred tank reactor (CSTR).

\par \textit{E. coli} growth was assumed to occur using acetic acid as the growth substrate \cite{Farmer1976} within a 2 L CSTR vessel. Constant operating conditions of 1.6 L broth volume and 37$^\circ$C were assumed. 

\par As part of the model, the growth of \textit{E. coli} was assumed to occur according to the irreversible reaction stoichiometry outlined in Equation \ref{eq: E.Coli growth stoichiometry}, following a literature value for the empirical composition of \textit{E. coli} \cite{bafna_ruhrer2024glucose}. 
\begin{equation}\label{eq: E.Coli growth stoichiometry}
    \begin{split}
        CH_3COOH + aO_2 + bNH_3 \rightarrow \\
        cCH_{1.8}O_{0.5}N_{0.2} + dCO_2 + eH_2O
    \end{split}
\end{equation}
\par The values of stoichiometric coefficients ($a$ to $e$) required were set to assume conservation of mass based on internal implementations within PyOMES.
\par It was also assumed that gas-liquid mass transfer and aqueous equilibria could occur within the vessel according to the stoichiometries previously outlined (Sections \ref{sec: pH prediction methods} and \ref{sec: gas-liquid methods}).   

\par As before, all aqueous equilibrium equations were assumed to occur instantaneously and
solved as  described in Section \ref{sec: pH prediction methods} and dynamic gas-liquid mass transfer was modelled as described in Section \ref{sec: gas-liquid methods}, with the same value of $k_La$ used. The rate of biomass growth and substrate depletion in the CSTR were modelled as occurring according to Monod-type kinetics (Equations \ref{eq: Monod biomass kinetics}-\ref{eq: Monod substrate kinetics}).
\begin{equation}\label{eq: Monod biomass kinetics}
    \frac{dX}{dt} = \left(\mu-D\right)X
\end{equation}
\begin{equation}\label{eq: Monod growth}
    \mu = \mu_{max,S}\left(\frac{S}{K_S + S}\right) \left(\frac{O_2}{K_{O_2}+O_2}\right)
\end{equation}
\begin{equation}\label{eq: Monod substrate kinetics}
    \frac{dS}{dt} = -\left(\frac{\mu}{Y_S}\right) X + D\left(S_{in}-S\right)
\end{equation}
where $X$ and $S$ are the concentrations of biomass and the carbon substrate in the liquid phase, respectively (g L$^{-1}$), $S_{in}$ is the substrate concentration in the inlet stream (g L$^{-1}$), $O_2$ is the dissolved oxygen concentration in the liquid phase (g L$^{-1}$), $\mu$ is the specific biomass growth rate (h$^{-1}$), $\mu_{\max,S}$ is the maximum specific growth rate under non-limiting substrate and oxygen conditions (h$^{-1}$), $K_S$ and $K_{O_2}$ are the substrate and dissolved oxygen half-saturation constants, respectively (g L$^{-1}$), $Y_S$ is the yield of biomass growth on the substrate (g$_X$ g$_S^{-1}$), and $D$ is the dilution rate (h$^{-1}$).

\par Values for $Y_S$, $\mu_{max,S}$, $K_S$ and $K_{O_2}$ were taken as 0.36 g$_X$ g$_S^{-1}$, 0.3 h$^{-1}$, and 0.5 g L$^{-1}$ and 0.2 mg $L^{-1}$ respectively.
a
\par The flowrate of liquid inlet and outlet were set equal at all times, maintaining a constant liquid volume. The dilution rate ($D$) was varied from 0.00 to 0.295 h$^{-1}$ as part of the investigations, with each simulation run to steady state. Steady state volumetric biomass productivity was calculated as the product of biomass concentration and dilution rate ($P=DX$). Steady state biomass and carbon substrate concentration were also calculated. The assumed composition of the CSTR feed stream is summarised in Table \ref{tab: CSTR compositions}. Beyond this, it was assumed that the composition of the vessel on start up was equal to that of the feed stream with the addition of a total of 0.05 g/L of \textit{E. coli}. 

\begin{table}[]
    \caption{Composition of material present in the reactor feed.}
    \centering
    \begin{tabular}{|l|l|}
        \hline
        Chemical        & Concentration, g/L \\
        \hline
        CH$_3$COOH      & 4.0                \\
        NH$_4$Cl        & 1.0                \\
        KH$_2$PO$_4$    & 3.0                \\
        \hline
    \end{tabular}
    \label{tab: CSTR compositions}
\end{table}

\section{Results and Discussion}
\subsection{Aqueous pH Prediction}\label{sec: aqueous pH results}
\par
\begin{figure}[h!]
    \centering
    \begin{subfigure}[b]{\linewidth}
        \centering
        \includegraphics[width=\linewidth]{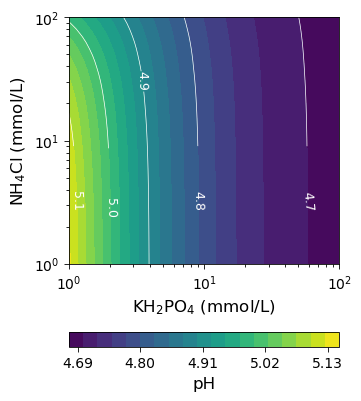}
        \label{fig:sub1}
    \vspace{.05cm}
    \end{subfigure}
    
    \begin{subfigure}[b]{\linewidth}
        \centering
        \includegraphics[width=\linewidth]{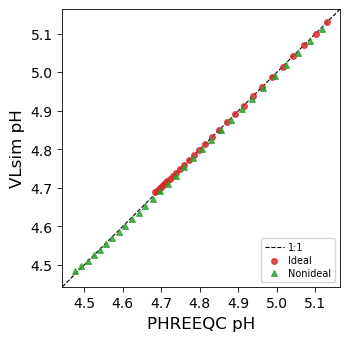}
        \label{fig:sub1}
    \end{subfigure}
    \caption{Model predictions for the aqueous pH prediction use case. The effect of solution concentrations on equilibrium pH according to the Davies activity model, simulated using PyOMES (top), and comparison of predictions made by PyOMES against PHREEQC \cite{parkhurst2013phreeqc} software as a benchmark (bottom). 
    }
    \label{fig: Aqueous pH prediction results}
\end{figure}

 Predicted pHs for the solutions are provided in Figure \ref{fig: Aqueous pH prediction results} alongside PHREEQC benchmark values. 
\par Results show PyOMES predicts the expected relationship between solution pH and the individual concentrations of KH$_2$PO$_4$ amd NH$_4$Cl. Additionally, curvature demonstrates the ability of the model to account for complex solution chemistry that may arise in systems of multiple equilibrium reactions as well as solution activity affects. 

\begin{table}[h!]
    \caption{Average run times required for aqueous pH prediction by simulation package  (n=500).}
    \centering
    \begin{tabular}{|l|l|l|}
        \hline
        Framework   & Run Time, ms \\
        \hline
        PyOMES      & 0.66              \\
        PHREEQC     & 1.27              \\ \hline
    \end{tabular}
    \label{tab: Aqueous pH Run Times}
\end{table}

\par Beyond trend predictions, the results in Figure \ref{fig: Aqueous pH prediction results} also show strong agreement between the prediction of PHREEQC software and PyOMES for both activity models considered. More specifically, maximum deviations of $<$0.001 are observed for predictions made by the PyOMES and PHREEQC models for the modelled solutions. Assuming correctness of the well-established PHREEQC software \cite{parkhurst2013phreeqc} implementation, this provides evidence for the reliability of the new PyOMES implementation. 

\par Finally, average run times for the calculation are provided in Table \ref{tab: Aqueous pH Run Times}. Results in Table \ref{tab: Aqueous pH Run Times} show an average run time of 0.66 ms per pH prediction made by PyOMES (depending on activity model). This is approximately 50\% faster than the average run times reported for PHREEQC (1.27 ms), suggesting PyOMES' performance is competitive relative to the benchmark. This suggests the possibility of running thousands of pH computations within seconds when using PyOMES - a throughput that can facilitate use of PyOMES for Monte Carlo methods, or within wider model-based research approaches. 


\subsection{Gas-Liquid Equilibration}\label{sec: gas-liquid results}
\par
\begin{figure}[h!]
    \centering
    \begin{subfigure}[b]{\linewidth}
        \centering
        \includegraphics[width=\linewidth]{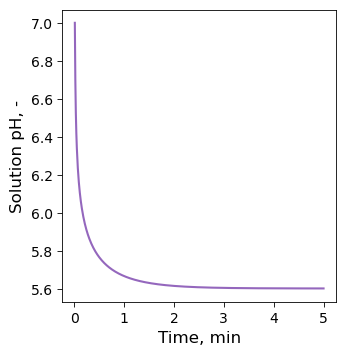}
        \label{fig:sub1}
    \vspace{.05cm}
    \end{subfigure}
    
    \begin{subfigure}[b]{\linewidth}
        \centering
        \includegraphics[width=\linewidth]{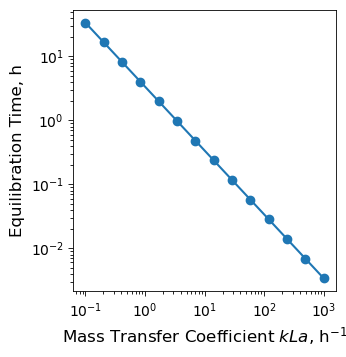}
        \label{fig:sub1}
    \end{subfigure}
    \caption{Model predictions for evolution of solution pH over time resulting from absorption of atmospheric CO$_2$ into pure water with $k_La = 96 h^{-1}$ (top) and predicted time taken to reach 95\% of the equilibrium value as a function of $k_La$ (bottom).
    }
    \label{fig: gas-liquid equilibration results}
\end{figure}

 A time series of the predicted solution pH for water in contact with air at atmospheric pressure and the relationship of the effective equilibration time to $k_La$ are shown in Figure \ref{fig: gas-liquid equilibration results}. 
 
\par Results in Figure \ref{fig: gas-liquid equilibration results} correctly predict the drop in solution pH over time for the modelled water solution that would be expected for the modelled scenario. This arises from the dissolution of CO$_2$ into water (and associated formation of aqueous carbonate species) which tends to have an acidifying effect on aqueous solutions. Importantly, the predicted pH value is found to converge to the expected theoretical value for the given conditions \cite{peart2000acid}.

\par By comparing results across a number of $k_La$ values it is also possible to see the importance of $k_La$ on the characteristic timescale taken for the modelled water-atmosphere system to reach effective equilibration. Specifically, results in Figure \ref{fig: gas-liquid equilibration results} show it is possible for the required equilibration time to span roughly four orders of magnitude over the $k_La$ values considered. This ranges from a time of minutes to days for $k_La$ values within the range of 0.1 to 100h$^{-1}$. This demonstrates how such trends predicted by PyOMES provide a powerful way of informing decision making during real experimental design. For example, this could allow researchers to identify solutions that are at risk of expiry if stored inadequately. 

\par Table \ref{tab: kLa computation times} reports the mean wall-clock time required to simulate one hour of gas--liquid equilibration. Across the investigated $k_La$ values, PyOMES required between 4.980 and 21.439 ms per simulated hour. These times are very reasonable for the possible use cases already described and provide capacity for running more advanced model-based investigations such as sensitivity and uncertainty analysis in tractable timescales. The runtime depends on the $k_La$ value as a result of use of an adaptive ODE solver, which employs shorter simulation timesteps for faster systems to maintain accuracy and stability.

\begin{table}[]
    \caption{Average wall-clock runtimes for gas-liquid equilibration simulations per simulated hour of experimental time (n=500).}
    \centering
    \begin{tabular}{|l|l|}
        \hline
        $k_La$ Value, h$^{-1}$    & Run Time, ms   \\ 
        \hline
        0.1                 & 4.980  \\
        10                  &  14.400\\
        1000                &  21.439\\
         \hline
    \end{tabular}
    \label{tab: kLa computation times}
\end{table}

\subsection{Aerobic Fermentation}
\begin{figure}[h!]
    \centering
    \begin{subfigure}[b]{\linewidth}
        \centering
        \includegraphics[width=\linewidth]{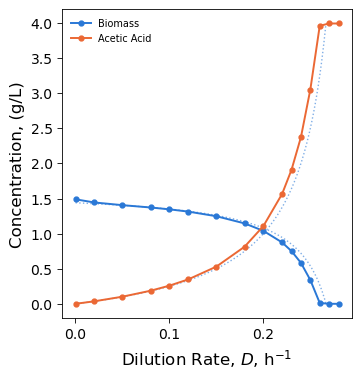}
        \label{fig:sub1}
    \vspace{.05cm}
    \end{subfigure}
    
    \begin{subfigure}[b]{\linewidth}
        \centering
        \includegraphics[width=\linewidth]{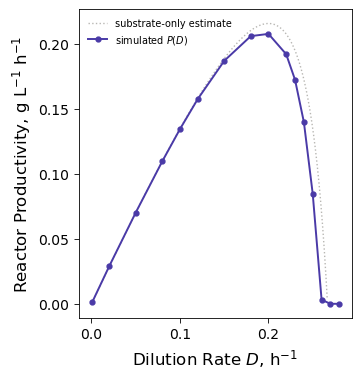}
        \label{fig:sub1}
    \end{subfigure}
    \caption{Model predictions for the steady state conditions of an aerobic fermentation CSTR as a function of dilution rate - biomass and acetic concentrations (top) and Reactor Productivity (bottom). Dotted lines reflect theoretical values for assuming a carbon substrate-limited chemostat CSTR.
    }
    \label{fig: CSTR results}
\end{figure}

Results in Figure \ref{fig: CSTR results} show predicted steady state behaviour as a function of operating dilution rate for the modelled CSTR. 

\par The results obtained are consistent with previously reported trends \cite{doran2013bioprocess}. Particularly, this includes the reduction in the concentration of biomass produced as dilution rate increases up to a value of 0.2 h$^{-1}$ followed by the onset of reactor washout shortly thereafter. This is combined with associated increase in concentration of acetic acid as biomass concentration is reduced. Similarly, the expected relationship of dilution rate and reactor productivity is achieved - indicating the onset of a maximum productivity at a critical dilution rate shortly before the onset of the washout condition. Importantly, this indicates the ability of PyOMES to reliably and easily implement more complex model formulations that may be of interest to a number of (bio)chemical processing fields. 
\par A summary of typical run times associated with CSTR simulations carried out in this section are provided in Table \ref{tab: CSTR computation times}. Results show the ability to handle the majority of CSTR simulations on the range of 1 to 10 s per run. Nonetheless, it is observed that simulation run time increases dramatically as the dilution rate approaches the washout condition for the reactor, with a dilution rate of 0.295 $h^{-1}$ requiring more than 1 minute of simulation time, compared to just 4 seconds with a dilution rate of 0.20 $h^{-1}$. This reflects the effect of dilution time on dynamics of the CSTR system. There is a particularly significant increase in the time required for the CSTR to reach steady-state as the dilution rate approaches the critical dilution rate, which is a well known process behaviour \cite{perram1973relaxation}. Reported computation times therefore suggest that PyOMES is able to  handle the CSTR simulation within reasonable timeframes for practical use cases. 

\begin{table}[]
    \centering
    \caption{Wall-clock runtimes for selected CSTR simulations from initialisation to steady-state convergence at different dilution rates (n=1).}
    \begin{tabular}{|c|c|}
        \hline
        Dilution rate   & Run Time, s \\
        \hline
         0.010           & 1.19 \\
         0.100           & 1.68 \\
         0.200           & 4.03 \\
         0.250           & 6.77\\
         0.290           & 41.76\\
         0.295           & 78.57 \\   
         \hline
    \end{tabular}
    \label{tab: CSTR computation times}
\end{table} 

\section{Conclusions and Future Work}
\subsection{PyOMES as a tool for (bio)chemical process simulations}
This work introduces the PyOMES package. The vision and architecture of the design laid out here may be understood as a foundation for understanding the package and its intended scope. The use cases considered in this report demonstrate that PyOMES is capable of modelling a broad range of (bio)chemical scenarios of interest to researchers and students working within the (bio)chemical sciences. Moreover, owing to the modular design of PyOMES, it is readily conceive how more complex modelling scenarios (than those shown here) could be established within the PyOMES modelling framework simply by repetition and expansion of the basic building blocks of the PyOMES architecture. This includes a number of features not explicitly shown in the uses cases such as the potential for multi-compartmental models, more sophisticated models for individual mass transport phenomena, and the ability to incorporate wider process behaviours such as control systems. Overall, it is therefore believed that PyOMES will find its place as a useful tool for the modelling and simulation of (bio)chemical process systems in future works. 

\subsection{User-Focused Continuous Development}
The modularity and approach to user-design taken by the PyOMES package make it accessible to researchers and students working in a broad range of possible (bio)chemical processing contexts. The ability to model such a broad range of processes under one software creates a great opportunity for a more unified, free, and open-source package to be established across fields. For this reason, it is expected that the future development of PyOMES has the potential to be directed by the efforts and interests of individual researchers using the platform. This may come about at least three main ways. Firstly, it may come from specific refinements to the canonical package itself based on input and troubleshooting raised by users. Secondly, it can come from user feedback and interest in implementation of well-known reference models within. Thirdly, it can come directly through the use and exploitation of the Protocol-led package architecture, which has been designed with the idea of enabling users to heavily customise the functionality of the package without the need for canonical updates to the package itself. 

\section{Package Availability}
The PyOMES package is available on GitHub: (\url{github.com/MGuo-Lab/PyOMES}) under a GNU Affero General Public License version 3 (AGPLv3) licence \cite{fsf_agpl3_2007}.

\section{Declarations}
\fontsize{11}{13}\selectfont

\textbf{Conflict of Interest: }
The authors declare no conflicts of interest.

\textbf{Funding Acknowledgments:}
The authors gratefully acknowledge funding received from the UK Department for Environment Food and Rural Affairs (DEFRA) through EU Green ERA-Hub under consortium project 'DAiry Waste and REsidues upCYCLing into Microbial ProtEin (DARE2CYCLE)'.TV and MG are grateful for the funding from Quorn and the BBSRC, under project reference 2725958. JL and MG would like to acknowledge financial support from the King's College London Net Zero Centre Ph.D.\ Scholarship scheme.

\textbf{Authors' Contributions:}
\par EE: Software Design \& Development, Conceptualisation, Methodology, Development \& Analysis of Use Cases, and Writing - Original Draft
\par TV: Software Design Discussion, Suggestion of Use Cases, Writing - Review \& Editing
\par JL: GUI Conceptualisation
\par MG: Conceptualisation, Supervision, Funding Acquisition

\printbibliography  

@book{Davies1962IonAssociation,
  author    = {Davies, C.W.},
  title     = {Ion Association},
  year      = {1962},
  publisher = {Butterworths},
  address   = {London}
}

@techreport{parkhurst2013phreeqc,
  title={Description of input and examples for {PhreeqQC} version 3 — {A computer program for speciation, batch-reaction, one-dimensional transport, and inverse geochemical calculations}},
  author={Parkhurst, D. L. and Appelo, CAJ},
  year={2013},
  institution={U.S. Geological Survey},
  number={6-A43}
}

@article{Helikar2021,
  title = {The Need for Research-Grade Systems Modeling Technologies for Life Science Education},
  DOI = {10.1016/j.molmed.2020.11.005},
  journal = {Trends in Molecular Medicine},
  publisher = {Elsevier BV},
  author = {Helikar,  Tomáš},
  year = {2021},
  month = Feb,
}

@article{Baio2016,
  title = {When Simple Becomes Complicated: Why Excel Should Lose its Place at the Top Table},
  DOI = {10.5301/grhta.5000247},
  journal = {Global and Regional Health Technology Assessment: Italian; Northern Europe and Spanish},
  publisher = {Aboutscience Srl},
  author = {Baio,  Gianluca and Heath,  Anna},
  year = {2016},
  month = Feb
}

@article{Djaffardjy2023,
  title = {Developing and reusing bioinformatics data analysis pipelines using scientific workflow systems},
  DOI = {10.1016/j.csbj.2023.03.003},
  journal = {Computational and Structural Biotechnology Journal},
  author = {Djaffardjy,  Marine and Marchment,  George and Sebe,  Clémence and Blanchet,  Raphaël and Belhajjame,  Khalid and Gaignard,  Alban and Lemoine,  Frédéric and Cohen-Boulakia,  Sarah},
  year = {2023},
  month = Jan,
}

@misc{fsf_agpl3_2007,
  author       = {{Free Software Foundation}},
  title        = {{GNU Affero General Public License, Version 3}},
  year         = {2007},
  month        = nov,
  day          = {19},
  howpublished = {\url{https://www.gnu.org/licenses/agpl-3.0.html}},
}

@inbook{GonzlezFigueredo2019,
  title = {Fermentation: Metabolism,  Kinetic Models,  and Bioprocessing},
  DOI = {10.5772/intechopen.82195},
  booktitle = {Current Topics in Biochemical Engineering},
  author = {González-Figueredo,  Carlos and Alejandro Flores-Estrella,  René and A. Rojas-Rejón,  Oscar},
  year = {2019},
  month = Aug 
}

@software{python310,
  author  = {{Python Software Foundation}},
  title   = {Python 3.10},
  version = {3.10},
  year    = {2021},
  url     = {https://www.python.org/downloads/release/python-3100/},
  note    = {Computer software}
}

@article{2020SciPy,
  author  = {Virtanen, Pauli and Gommers, Ralf and Oliphant, Travis E. and
             Haberland, Matt and Reddy, Tyler and Cournapeau, David and
             Burovski, Evgeni and Peterson, Pearu and Weckesser, Warren and
             Bright, Jonathan and {van der Walt}, St{\'e}fan J. and
             Brett, Matthew and Wilson, Joshua and Millman, K. Jarrod and
             Mayorov, Nikolay and Nelson, Andrew R. J. and Jones, Eric and
             Kern, Robert and Larson, Eric and Carey, C J and
             Polat, {\.I}lhan and Feng, Yu and Moore, Eric W. and
             {VanderPlas}, Jake and Laxalde, Denis and Perktold, Josef and
             Cimrman, Robert and Henriksen, Ian and Quintero, E. A. and
             Harris, Charles R. and Archibald, Anne M. and
             Ribeiro, Ant{\^o}nio H. and Pedregosa, Fabian and
             {van Mulbregt}, Paul and {SciPy 1.0 Contributors}},
  title   = {{{SciPy} 1.0: Fundamental Algorithms for Scientific
             Computing in Python}},
  journal = {Nature Methods},
  year    = {2020},
  volume  = {17},
  number  = {3},
  pages   = {261--272},
  doi     = {10.1038/s41592-019-0686-2}
}

@article{Farmer1976,
  title = {The Energetics of Escherichia coli during Aerobic Growth in Continuous Culture},
  DOI = {10.1111/j.1432-1033.1976.tb10639.x},
  journal = {European Journal of Biochemistry},
  author = {FARMER,  Ian S. and JONES,  Colin W.},
  year = {1976},
  month = Aug,
}

@book{Rawlings2026ModelPredictiveControl,
  author    = {Rawlings, J. B. and Mayne, D. Q. and Diehl, Moritz M.},
  title     = {Model Predictive Control: Theory, Computation, and Design},
  year      = {2026},
  publisher = {Nob Hill Publishing},
}

@article{Rebello2025DigitalTwins,
  author  = {Rebello, C.M. and Nogueira, I.B.R.},
  title   = {Digital twins in chemical engineering: An integrated framework for identification, implementation, online learning, and uncertainty assessment},
  journal = {Computers \& Chemical Engineering},
  year    = {2025},
  doi     = {10.1016/j.compchemeng.2025.109178},
}

@article{Vinestock2026FusariumFermentation,
  author    = {Vinestock, T. and Guo, M.},
  title     = {Parameter Estimation and Model Comparison for Mixed 
                Substrate Biomass Fermentation},
  journal   = {Systems and Control Transactions},
  year      = {2026},
  publisher = {PSE Press},
  doi       = {10.69997/sct.178293},
}

@article{Geremia2026ModelBasedDesign,
  author  = {Geremia, M. and Macchietto, S. and Bezzo, F.},
  title   = {A review on model-based design of experiments for parameter precision – Open       challenges, trends and future perspectives},
  journal = {Chemical Engineering Science},
  year    = {2026},
  doi     = {10.1016/j.ces.2025.122347},
}

@software{heinsbroek_phreeqpython,
  author  = {Heinsbroek, Abel},
  title   = {{PhreeqPython}: An Object-Oriented Python Wrapper for the
             {VIPhreeqc} Module},
  year    = {2026},
  version = {1.6.2},
  url     = {https://github.com/Vitens/phreeqpython},
}

@article{bafna_ruhrer2024glucose,
  author  = {Bafna-R{\"u}hrer, Jonas and Bhutada, Yashomangalam D. and
             Orth, Jean V. and {\O}zmerih, S{\"u}leyman and Yang, Lei and
             Zielinski, Daniel and Sudarsan, Suresh},
  title   = {Repeated glucose oscillations in high cell-density cultures
             influence stress-related functions of {Escherichia coli}},
  journal = {PNAS Nexus},
  year    = {2024},
  doi     = {10.1093/pnasnexus/pgae376}
}

@book{doran2013bioprocess,
  author    = {Doran, Pauline M.},
  title     = {Bioprocess Engineering Principles},
  edition   = {2},
  year      = {2013},
  publisher = {Academic Press},
  isbn      = {978-0-12-220851-5}
}

@article{perram1973relaxation,
  author  = {Perram, John W.},
  title   = {Relaxation times in bacteriological culture and the approach to steady state},
  journal = {Journal of Theoretical Biology},
  volume  = {38},
  number  = {3},
  pages   = {571--578},
  year    = {1973},
  doi     = {10.1016/0022-5193(73)90257-9}
}

@article{peart2000acid,
  author  = {Peart, M. R.},
  title   = {Acid Rain, Storm Period Chemistry and Their Potential
             Impact on Stream Communities in {Hong Kong}},
  journal = {Chemosphere},
  year    = {2000},
  doi     = {10.1016/S0045-6535(99)00386-0}
}

\newpage
\onecolumn
\title{\fontsize{16}{20}\selectfont\textbf{Supporting Information}}
\section{Solution of Instantaneous Mass Transport Equations in PyOMES}
\par Instantaneous mass transport equilibrium (aqueous chemical reaction equilibria and physical gas-liquid equilibria) were calculated by solving the coupled mass-balance and charge-balance equations together with the equilibrium relationships expressed in terms of the logarithms of the equilibrium constants, ($\log K$), and species activities. A summary of log$K$ values assumed for each aqueous equilibrium reaction considered within models is provided in Tables \ref{tab: SI aqueous logk values} and \ref{tab: SI henry coefficients}.

\par Solution of system of equations produced by this formulation was handled by a PyOMES-native solver. This solver uses the Newton-Raphson method to find solutions based on numerical root finding. For further information refer to the PyOMES source code.

\begin{table}[h!]
    \caption{Thermodynamic data associated with aqueous chemical reaction equilibria modeled in the use cases of this work.}
    \centering
    \begin{tabular}{| l | l | l |} \hline
        Reaction & $logK$ & Description \\ \hline
        $H_2O(l) \rightleftharpoons H^+(aq) + OH^-(aq)$ & -14.0  & water autoionisation \\
         
        $H_3PO_4(aq) \rightleftharpoons H_2PO4^-(aq) + H^+(aq)$ & -2.15 & phosphate dissociation 1 \\
         
        $H_2PO4^-(aq) \rightleftharpoons  HPO4^{2-}(aq) + H^+(aq)$ & -7.20 & phosphate dissociation 2 \\
         
        $HPO4^{2-}(aq) \rightleftharpoons PO4^{3-}(aq) + H^+(aq)$ & -12.35  & phosphate dissociation 3 \\
         
        $NH_4^+(aq)\rightleftharpoons NH_3(aq) + H^+(aq)$ & -9.25 & ammonium dissociation \\
         
        $CO_2(aq) + H_2O(l) \rightleftharpoons HCO_3^{-}(aq) + H^{+}(aq)$ & -6.35 & apparent carbonate dissociation 1 \\

        $HCO_3^-(aq) \rightleftharpoons CO_3^{2-}(aq) + H^+(aq)$ & -10.33 & bicarbonate dissociation \\

        $CH_3COOH(aq) \rightleftharpoons CH_3COO^-(aq) + H^+(aq)$ & -4.76 & acetate dissociation \\
        \hline
    \end{tabular}
    \label{tab: SI aqueous logk values}
\end{table}

\begin{table}[h!]
    \caption{Thermodynamic data associated with gas-liquid equilibria modeled in the use cases of this work.}
    \centering
    \begin{tabular}{| l | l | l |} \hline
        Reaction & $H_{ref}$, mol m$-^3$ Pa$^{-1}$ & $\mathrm{d}\ln(H)$,  \\ 
        \hline
        $H_2(g) \rightleftharpoons H_2(aq)$ & 3.4x10$^{-4}$  & 2400 \\
         
        $CO_2(g) \rightleftharpoons CO_2(aq)$ & 1.3x10$^{-5}$ & 1500 \\
         
        $N_2(g) \rightleftharpoons  N_2(aq)$ & 6.4x10$^{-6}$ & 1600 \\
        \hline
    \end{tabular}
    \label{tab: SI henry coefficients}
\end{table}

\begin{equation}\label{eq: logK equation}
    K_r = \prod_{i} (\gamma_i C_i)^{v_{i,r}}
\end{equation}
where $K_r$ is the equilibrium constant for reaction $r$, $\gamma_i$ is the activity coefficient for species $i$ at equilibrium, $C_i$ is the solution concentration of target species $i$, and $v_i$ is the stoichiometric coefficient associated with species $i$ present within the associated reaction $r$ with convention that $v_i$ is negative for reactants and positive for products. 


\begin{equation}\label{eq: Henry equation}
    H_i = \frac{C_i}{P_i}
\end{equation}
where $H_i$ is the Henry constant relating the equilibrium partial pressure $P_i$ (Pa) and solution concentration $C_i$ (mol m$^-3]$) for compound $i$, 

\subsection{Use of the PHREEQC software}
The PHREEQC software \cite{parkhurst2013phreeqc} was selected as an alternative software implementation with which to benchmark the prediction accuracy and computational speed of the PyOMES package against. This was done using the phreeqpython python package \cite{heinsbroek_phreeqpython} using the standard thermodynamic dataset provided within the PHREEQC vitens.dat dataset. This dataset relies on the the WATEQ Debye-Hückel activity correction native to the PHREEQC software (see Equation \ref{eq: WATEQ equation}. The WATEQ model differs slightly from the Davies equation (Equation \ref{eq: Davies equation}) applied in the PyOMES software, though predictions were found empirically to be within good agreement for the use cases studied in this work. For further information, refer to \cite{parkhurst2013phreeqc}.

\begin{equation}
    \log_{10}\gamma_i
    =
    -\frac{A z_i^2 \sqrt{I}}
    {1+B a_i^{\circ}\sqrt{I}}
    + b_i I,
    \label{eq: WATEQ equation}
\end{equation}

\begin{equation}
    \log_{10}\gamma_i
    =
    -A z_i^2
    \left(
        \frac{\sqrt{I}}{1+\sqrt{I}}
        - 0.3I
    \right),
    \label{eq: Davies equation}
\end{equation}

\end{document}